\documentclass[journal]{IEEEtran}
\ifCLASSINFOpdf
\else
\fi
	\usepackage[comma,numbers,square,sort&compress]{natbib}
	\usepackage{algorithm} 
	\usepackage{algorithmic} 

	\usepackage{multirow} 
	\usepackage{xcolor}
	\usepackage{setspace}
	\usepackage{amsmath}
	\newtheorem{theorem}{\textbf{Theorem}}
	\newtheorem{lemma}{\textbf{Lemma}}

	\newtheorem{remark}{\textbf{Remark}}
	\newtheorem{definition}{\textbf{Definition}}
	\newtheorem{problem}{\textbf{Problem}}
	\newtheorem{proposition}{\textbf{Proposition}}
	\newtheorem{assumption}{\textbf{Assumption}}

		\newenvironment{proof}{{\noindent{\bf \emph{Proof:}}}\quad}{\hfill $\square$\par}
		
		\usepackage{multirow}
		\usepackage{url}
		\usepackage{subfigure}
		\usepackage{fancyhdr}
		\usepackage{amsmath}
		\usepackage{multirow}
		\usepackage{amssymb }
		\usepackage{color}
		\usepackage{graphics} 
		\usepackage{graphicx}
		\usepackage{geometry}

\begin{document}
			%
				%
				%
			%
			
			\title{\LARGE \bf
				Observability Blocking for Functional Privacy of \\~ Linear Dynamic Networks
			}
			
			%
			%
			
			\author{Yuan Zhang, Ranbo Cheng, and Yuanqing Xia
				\thanks{The authors are with the School of Automation, Beijing Institute of Technology, Beijing, China. 
					{(email: {\tt\small zhangyuan14@bit.edu.cn, chengranbo123@163.com, xia\_yuanqing@bit.edu.cn}).} }}
\maketitle


\begin{abstract}
	This paper addresses the problem of determining the minimum set of state variables in a network that need to be blocked from direct measurements in order to protect functional privacy with respect to {\emph{any}} output matrices. The goal is to prevent adversarial observers or eavesdroppers from inferring a linear functional of states, either vector-wise or entry-wise. We prove that both problems are NP-hard. However, by assuming a reasonable constant bound on the geometric multiplicities of the system's eigenvalues, we present an exact algorithm with polynomial time complexity for the vector-wise functional privacy protection problem. Based on this algorithm, we then provide a greedy algorithm for the entry-wise privacy protection problem. Our approach is based on relating these problems to functional observability and leveraging a PBH-like criterion for functional observability. Finally, we provide an example to demonstrate the effectiveness of our proposed approach.

\end{abstract}
\begin{IEEEkeywords}
	Observability blocking, network privacy preservation, functional observability, algorithms
\end{IEEEkeywords}


\section{Introduction}
In recent years, research on the privacy preservation of control systems has gained increasing attention due to the growing use of networked control systems, cyber-physical systems and the increasing concerns over the privacy and security of these systems \cite{ding2018survey,darup2021encrypted,xia2022brief}.

One area of research has focused on developing privacy-preserving control algorithms that can achieve the desired control objectives while protecting sensitive information about the system's states and inputs \cite{mo2016privacy,wang2019privacy}. Another area of research has explored the use of differential privacy techniques to protect the privacy of data collected from control systems \cite{han2016differentially,cortes2016differential}. Recently, encryption techniques, including homomorphic and nonhomomorphic encryptions, have also been adopted to preserve data privacy in the process of network transmissions and third-party computations \cite{lu2018privacy,darup2021encrypted,xia2022brief}. Additionally, researchers have also investigated the impact of cyber-attacks on the privacy and proposed secure and privacy-preserving communication protocols to mitigate these attacks \cite{keshk2019integrated}.

Apart from the above research perspectives, there has been a natural relationship between the privacy preservation of control systems and system observability \cite{pequito2014design,al2019observability,lu2020privacy,zhang2018observability}. Observability refers to the ability to estimate the state of a system based on its output. In the context of privacy preservation, the idea is to make it difficult for an adversary to infer sensitive information about the system by limiting their ability to observe the system's state. In this line, privacy preservation through system design has attracted much interest. To name a few, \cite{pequito2014design} considered the problem of designing communication networks so that the average consensus is achieved while the observable subspace each individual agent can infer from shared information of its neighbors is as small as possible.  The design of local state-feedback control systems in dynamical networks to block observability at remote nodes was studied in \cite{al2019observability}. Leveraging the notion of non-strong observability, \cite{lu2020privacy} considered adding perturbations to system inputs and outputs to protect partial entries of the initial states and inputs. The problem of blocking a minimum set of state variables from being measured by existing sensors to destroy observability was considered in \cite{zhang2018observability}. The basic idea in these works to achieve privacy preservation is to design control systems with reduced observability. This means that the system state cannot be easily inferred from its output, making it more difficult for an adversary to infer sensitive information.

In this paper, we take a step further in the direction of protecting network privacy by considering functional privacy, i.e., linear functionals of states that need to be kept confidential to adversarial observers. The goal is to identify the minimum set of state variables (nodes) in a network that need to be blocked from direct measurements to prevent the inference of a given functional privacy with respect to {\emph{any}} output matrices. We consider two different privacy protection levels: vector-wise protection and entry-wise protection, meaning that the functional privacy in the vector form cannot be inferred as a whole and that every component (entry) of it cannot be inferred, respectively.

By relating functional privacy protection to the notion of functional observability and leveraging a PBH-like criterion for functional observability \cite{fernando2010functional,jennings2011existence,fernando2010functional2}, we make the following contributions. First, we prove that both functional privacy protection problems are NP-hard. Second, assuming a reasonable constant bound on the geometric multiplicities of the system's eigenvalues, we provide an exact algorithm with polynomial time complexity for the vector-wise problem. Third, we provide a greedy algorithm for the entry-wise problem. Our results reveal the role of node measurements in protecting the functional privacy of linear dynamic networks and enable us to identify which set of nodes can be protected at a lower cost to preserve functional privacy more efficiently. These nodes can be regarded as ``critical nodes'' that may leak confidential information and require specific protection measures.

The rest is organized as follows. Section \ref{section_prob} presents the problem formulation, and Section \ref{sec-III} provides preliminaries on functional observability. The complexity of the considered problems is given in the next section. Section \ref{sec-V} presents algorithms for these problems, followed by an illustrative example in Section \ref{sec-VI}. The last section concludes this paper.

{\emph{Notations:}} For a set, $|\cdot |$ denotes its cardinality. The symbol $[n]=\{1,2,...,n\}$. A matrix $L$ is also denoted by $L=[l_{ij}]$ or $L=[L_{ij}]$, which means $l_{ij}$ or $L_{ij}$ is the entry in the $i$th row and $j$th column of $L$. By ${\rm eig}(M)$ we denote the set of eigenvalues of the square matrix $M$. Let ${\bf diag}\{X_i|_{i=1}^n\}$ be the block diagonal matrix whose $i$th diagonal block is $X_i$, and ${\bf col}\{X_i|_{i=1}^n\}$ be the matrix stacked by $X_i|_{i=1}^n$. $I_n$ denotes the $n$ dimensional identify matrix, where the subscript $n$ may be omitted if it can be inferred from the context. By $e_i$ we denote the $i$th column of $I_n$, and ${\bf 1}_{m\times n}$ the $m\times n$ matrix with entries all being one. Given an $m\times n$ matrix $M$ and a set ${\cal S}\subseteq [n]$, ${M}_{\cal S}$ denotes the sub-matrix of $M$ formed by {\emph{rows}} indexed by ${\cal S}$, and ${M}^{\cal S}$ denotes the matrix obtained from $M$ by preserving its columns indexed with ${\cal S}$ and zeroing the rest. If $M=I$, $M_{\cal S}$ and $M^{\cal S}$ coincide.

\section{Problem Formulation} \label{section_prob}
Consider a network of $n$ nodes. The $i$th node evolves according to the following dynamics
\begin{equation} \label{sub_dynamic}
	\dot x_i(t)= a_{ii}x_i(t) + \sum_{j\in [n]\backslash \{i\}} a_{ij}x_j(t)+\sum_{j=1}^mb_{ij}u_{j}(t),
\end{equation}
where $x_i(t)\in {\mathbb R}$ is the state variable of the $i$th node, $u_j(t)\in {\mathbb R}$ is the $j$th input, $a_{ii}\in {\mathbb R}$ is the self-damping coefficient, $a_{ij}\in {\mathbb R}$ is the coupling strength from node $j$ to node $i$, and $b_{ij}\in {\mathbb R}$ stands for the affection from the $j$th input to the $i$th node. The topology of this network can be described by a directed graph ${\cal G}=({\cal V},{\cal E})$, where ${\cal V}=\{1,...,n\}$ denotes the node set and ${\cal E}=\{(i,j): a_{ji}\ne 0\}$ denotes the edge set. Let $x(t)=[x_1(t),\cdots, x_n(t)]^\intercal$, $u(t)=[u_1(t),\cdots,u_m(t)]^\intercal$, $A=[a_{ij}]$ and $B=[b_{ij}]$. The lumped form of $(\ref{sub_dynamic})$ can be written as
\begin{equation} \label{lumped-system}
	\dot x(t)=Ax(t)+Bu(t).
\end{equation}
System (\ref{sub_dynamic}) can describe the dynamics of a linear time-invariant network system. A typical example is the multi-agent system where each agent is a single integrator running the consensus protocol \cite{mesbahi2010graph}. In this case, $a_{ii}=-\sum_{j\in [n]\backslash \{i\}}a_{ij}$, meaning that $-A$ is the weighted Laplacian matrix of ${\cal G}$. Other examples include interacted liquid tanks \cite{zhang2021structural}, synchronizing networks of linear oscillators \cite{LucaSynchronization}, opinion propagation in social networks \cite{ye2018evolution}, etc.

The output of system (\ref{lumped-system}) is given by
\begin{equation} \label{lumped-output}
	y(t)=Cx(t),
\end{equation}with $C\in {\mathbb R}^{p\times n}$ the output matrix. Let $y(x,u,t)$ be the output signal of system (\ref{lumped-system})-(\ref{lumped-system}) at time $t$ generated from the
initial state $x$ by the input $u$.  Each nonzero row of $C$ corresponds to a sensor. If there is only one nonzero entry in a row of $C$, we call the sensor associated with this row a dedicated sensor, meaning that this sensor measures only one state variable.

Suppose an adversarial observer or eavesdropper intends to infer the information
\begin{equation} \label{functional-info}
	z(t)=Fx(t),
\end{equation}
where $F\in {\mathbb R}^{r\times n}$. Write $F$ as $F={\bf col}\{f_i|_{i=1}^r\}$, $f_i\in {\mathbb{R}^{1\times n}}$. As commonly assumed, the observer knows system parameters $(A,C,F)$ and has access to the signals $u(t)$ and $y(t)$ for a sufficiently long time horizon \cite{mo2016privacy,lu2020privacy,pequito2014design}.  The vector $z(t)$ can be some private or confidential information that the network wants to protect from being inferred by the observer, and is called {\emph{functional privacy}} since it is a linear combination of the state $x(t)$. Typical examples of $z(t)$ include:
\begin{itemize}
	\item  the full state $x(t)$ when $F=I_n$;
	\item the vector formed by a set of target states indexed by ${\cal S}\subseteq [n]$ when $F={\bf col}\{e_i^\intercal|_{i\in {\cal S}}\}$;
	\item  the average of all states when $F=\frac{1}{n}{\bf 1}_{1\times n}$;
	\item the vector consisting of averages of states of clusters indexed by ${\cal S}_1,...,{\cal S}_l\subseteq [n]$ when $F={\bf col}\{\frac{1}{|{\cal S}_i|}{\bf 1}^{{\cal S}_i}|_{i=1}^l\}$. \end{itemize} We remark that designing observers to infer the average state or average cluster states has been considered in \cite{niazi2019scale,niazi2020average}.

There are typically several ways to protect the functional privacy $z(t)$, for example, adding noise to the output $y(t)$ in the spirit of differential privacy \cite{le2013differentially}, or using certain encryption techniques to encrypt $y(t)$ \cite{lu2018privacy}. In this paper, however, we consider the structure requirement on the output matrix $C$, under which adversarial observers cannot infer the functional privacy $z(t)$. The advantage of doing so is that by designing an appropriate measurement structure we can preserve the functional privacy {\emph{without}} using any privacy-preserving techniques.

\begin{definition}[Vector-wise functional privacy protection]\label{def-inferable} The functional privacy $z(t)=Fx(t)$ is inferable, if for any initial state $x(0)$ and input $u(t)$, there exists a finite time $t_f$ such that the initial value of the function $Fx(0)$ can be uniquely determined from the observation $y(t)$ and input $u(t)$, $0\le t \le t_f$.  The functional privacy $z(t)$ is said to be (vector-wisely) protected if it is not inferable. 		
\end{definition}

\begin{definition}[Entry-wise functional privacy protection] The functional privacy $z(t)=Fx(t)$ is said to be entry-wisely protected if every component of $z(t)$, i.e., $z_i(t)=f_ix(t)$, is protected for $i=1,...,r$.		
\end{definition}
\begin{remark}
	The case that $z(t)$ is not inferable does not imply that every component of it is not inferable. There may exist scenarios where $z(t)$ is not inferable but its partial components are. Therefore, the entry-wise protection is stricter than the vector-wise protection. From Definition \ref{def-inferable},  with the knowledge of $(A,C,F)$ and $u(t),y(t)$, $t\in [0,t_f]$, if an observer can infer $z(0)$, then it can infer $z(t)$ for $t\in [0,t_f]$.
\end{remark}

In the network context, the structure of $C$ can be dominantly determined by the state variables (nodes) it directly measures.  We say a set of state variables indexed by ${\cal S}\subseteq [n]$ is blocked from direct measurement with respect to the output matrix $C$ (blocked w.r.t. $C$ for short) if the columns of $C$ indexed by ${\cal S}$ are turned to zeros (i.e., $C$ is turned to $C^{[n]\backslash {\cal S}}$).  To see how many state variables should be blocked w.r.t. whatever output matrices $C$ to protect the functional privacy, we consider the following two problems:

\begin{problem} \label{prob1}
	How can we select the minimum set of state variables to be blocked w.r.t. {\emph{any}} output matrix $C$ such that $z(t)$ is protected? \end{problem}

\begin{problem} \label{prob2}
	How can we select the minimum set of state variables to be blocked w.r.t. {\emph{any}} output matrix $C$ such that {\emph{every}} component of $z(t)$ is protected? \end{problem}

A trivial solution to Problems 1 and 2 is the full state set, under which $y(t)$ cannot convey any information of $z(t)$. This implies these two problems are well-defined.
Problems 1 and 2 enable a better understanding of the role of nodal measurements in the preservation of the given functional privacy $z(t)$. Note the problems do not depend on a specific output matrix $C$, meaning that the solutions are properties of the network system matrix $A$ and the functional privacy characterized by $F$. The solutions to these problems can somehow tell us the set of nodes that we can protect with less cost to preserve functional privacy more efficiently (since blocking more nodes from direct measurement means that more energy or additional effort is needed). System (\ref{lumped-system})-(\ref{functional-info}) may be represented by the pair $(A, F)$ when Problems 1 and 2 are considered.  	

It is obvious that in Problems 1 and 2, it suffices to consider the case with dedicated sensors, i.e., $C=I_n$. This is because, for any output matrix $C\in {\mathbb R}^{m\times n}$ and a given ${\cal S}\subseteq [n]$, even $m>n$, upon defining $\bar {\cal S}=[n]\backslash {\cal S}$, the new output matrix $C^{\bar {\cal S}}$ after blocking state variables ${\cal S}$ means that at most $n-|{\cal S}|$ state variables are measured so that we can extract no more than $n-|{\cal S}|$ individual states indexed by $\bar {\cal S}$.




\section{Functional observability} \label{sec-III}
In this section, we investigate Problems 1 and 2 in the spirit of functional observability, resulting in reformulations for them.

We shall denote the system described by (\ref{lumped-system})-(\ref{functional-info}) as the triple $(A,C,F)$.	Before introducing the definition of functional observability, we give an equivalent definition of being inferable as follows.

\begin{lemma}\label{def-inferable-2} The functional privacy $z(t)=Fx(t)$ is inferable (by an adversarial observer), if and only if for any initial states $x_1$ and $x_2$ and input $u$, $y(x_1,u,t)=y(x_2,u,t)$ for all $t\ge 0$ implies that $Fx_1=Fx_2$.
\end{lemma}

\begin{proof}
	The sufficiency is obvious since this condition implies for any output $y(t)$ and input $u(t)$, there is a unique initial function $Fx(0)$ that obeys the system dynamics. For the necessity, suppose there exist two initial states $x_1,x_2$ and input $u$ such that $y(x_1,u,t)=y(x_2,u,t)$ for all $t\ge 0$ but $Fx_1\ne Fx_2$. Then, the initial function $Fx(0)$ cannot be determined uniquely.
\end{proof}

\begin{definition}[Observability]\cite{H.Tr2012.Control}
	System (\ref{lumped-system})-(\ref{lumped-output}) is said to be observable, if for any initial states $x_1,x_2$ and the zero input, $y(x_1,0,t)=y(x_2,0,t)$ for all $t\ge 0$ implies that $x_1=x_2$.
\end{definition}

\begin{definition}[Functional observability]\cite{fernando2010functional2,jennings2011existence} \label{functional-def}
	System (\ref{lumped-system})-(\ref{functional-info}) is said to be functionally observable, if for any initial states $x_1,x_2$ and input $u$, $y(x_1,u,t)=y(x_2,u,t)$ for all $t\ge 0$ implies that $Fx_1=Fx_2$.
\end{definition}

In other words, functional observability is the ability to infer linear functions of states $Fx(t)$ from the knowledge of external inputs $u(t)$ and outputs $y(t)$ of a system \cite{fernando2010functional}. When $F=I_n$, functional observability collapses to conventional observability. Based on functional observability and Lemma \ref{def-inferable-2}, the functional privacy $z(t)=Fx(t)$ is inferable for a system $(A,C,F)$, if and only if $(A,C,F)$ is functionally observable. In addition, $Fx(t)$ is entry-wisely protected, if and only if $(A,C,f_i)$ is not functionally observable for $i=1,...,r$, recalling $f_i$ is the $i$th row of $F$. Therefore, Problems 1 and 2 can be equivalently formulated as
\[ \begin{array}{*{20}{l}}
	{\bf {Problem \ 1:}} \\
	\ \ \ \ 	{\mathop {\min }\nolimits_{{\cal S} \subseteq [n]} {\kern 1pt} {\kern 1pt} {\kern 1pt} {\kern 1pt} {\kern 1pt} \left| {\cal S}  \right|}  \\
	\ \ \ \ 	{{\rm{s.t. }}\ (A,{I_{[n]\backslash {\cal S}}},F){\kern 1pt}\ {\rm{not}} {\kern 1pt}\ {\rm{functionally}} {\kern 1pt}\ {\rm{observable.}}} \\
	{\bf {Problem \ 2:}}\\
	\ \ \ \ 	{\mathop {\min }\nolimits_{{\cal S} \subseteq [n]} {\kern 1pt} {\kern 1pt} {\kern 1pt} {\kern 1pt} {\kern 1pt} \left| {\cal S}  \right|}\\
	\ \ \ \ 	{{\rm{s.t.}}\ (A,{I_{[n]\backslash S}},f_i){\kern 1pt}\ \ {\rm{not\ functionally\ observable,}} \ \forall i\in [r].}
\end{array}\]

The following lemma is a revised version of \citep[Theo. 4]{jennings2011existence}, \citep[Theo.2]{fernando2010functional2}, and \citep[Lem. 2.3]{fernando2014system},  which gives a necessary and sufficient condition for functional observability under the diagonalization assumption on $A$. {\emph{This revision is of independent interest of this paper and is scheduled to be elaborated separately.}}   

\begin{lemma}\label{functional-theorem}[Revising \citep[Theo. 4]{jennings2011existence}] Suppose that $A$ is diagonalizable. The triple $(A,C,F)$ is functionally observable if and only if
	\begin{equation}\label{rank-criterion} {\rm rank}  \left[\begin{array}{c}
			A-\lambda I_n \\
			C \\
			F
		\end{array}\right]={\rm rank}  \left[\begin{array}{c}
			A-\lambda I_n \\
			C
		\end{array}\right], \forall \lambda\in {\mathbb C}.\end{equation}	
\end{lemma}

It can be seen that when $F=I_n$, the above condition collapses to the PBH test for the conventional observability.  When $(A,C,F)$ is functionally observable, one can find a matrix $F_0$ satisfying two additional conditions (see \citep[Theo. 2]{fernando2010functional}), based on which a functional observer with arbitrary poles can be constructed to estimate $z(t)$ asymptotically; see \cite{{fernando2010functional}} for details.

\section{Complexity Analysis} \label{sec-IV}
In this section, we prove that both Problems 1 and 2 are NP-hard.
When $F$ is a row vector, Problem 1 reduces to Problem 2. This indicates to show the NP-hardness of Problem 1, it suffices to show the NP-hardness of Problem 2 with a scalar functional privacy. To proceed with our proof, the following lemmas are needed.

\begin{lemma}\citep[Chap. 0.2.7]{Horn2013Matrix}\label{composity}
	Let $M={\bf col}\{M_1,M_2\}$ be a composite matrix and $M_1^\bot$ (if exists) consist of a set of linearly independent column vectors spanning the null space of $M_1$ ($M_1^\bot$ is called a basis matrix). Then, $M$ is of full column rank, if and only if $M_2M_1^\bot$ is of full column rank.
\end{lemma}

\begin{lemma}[Cauchy bound]\cite{hirst1997bounding} \label{Cauchy-bound} All roots of the polynomial $f(z)=a_nz^n+a_{n-1}z^{n-1}+\cdots+a_1z+a_0$ with $a_n\ne 0$ lie in the disk
	$$|z|<1+\max_{0\le k\le n-1} |a_k/a_n|.$$
\end{lemma}

\begin{lemma}\label{f-construct} Let $W={\bf col}\{w_i|_{i=1}^n\}$ be an $n\times k$ ($k<n$) integer matrix with full column rank,  where $w_i\in {\mathbb R}^{1\times k}$ is a row vector with integral entries. Moreover,
	let $\beta_{\max}=\max \{|w_{ij}|\}$, in which $w_{ij}$ is the $(i,j)$th entry of $W$. Then, for any $\alpha\ge 1+k^k\beta^k_{\max}$, and any ${\cal S}\subseteq [n]$ with $|{\cal S}|=k-1$ such that ${\rm rank }W_{{\cal S}}=k-1$, the following matrix has full column rank
	$$W(\alpha)\doteq \left[\begin{array}{c}
		\alpha w_1+\alpha^2w_2+\cdots+ \alpha^nw_n \\
		W_{{\cal S}}
	\end{array}\right].$$			
\end{lemma}
\begin{proof}
	By the property of determinant,
	$$\det W(\alpha)=\sum\nolimits_{i=1}^n \alpha^i\det W_{\{i\}\cup {\cal S}},$$
	where $W_{\{i\}\cup {\cal S}}\doteq{\bf col}\{w_i,W_{\cal S}\}$. Since $W$ has full column rank, for any ${\cal S}$ with the property mentioned above, there exists $j\in [n]\backslash {\cal S}$ such that $W_{\{j\}\cup {\cal S}}$ is non-singular. This means, at least one item, say $\alpha^{j}$, has a nonzero coefficient $\det W_{\{j\}\cup {\cal S}}$ in $\det W(\alpha)$.
	Therefore, $\det W(\alpha)$ cannot be identically zero. Notice that the determinant of any $k\times k$ submatrix of $W$ is no more than $\beta_{\max}^kk!\le \beta^k_{\max}k^k$. This is due to the fact that the determinant of a $k\times k$ matrix is the sum of signed
	products of all $k$ entries with the property that any two of them come from different rows and columns of this matrix. Consequently, all coefficents of $\alpha^i$, $1\le i\le n$, have absolute values bounded by $k^k\beta_{\max}^k$. Since the polynomial $\det W(\alpha)$ is not identically zero and the leading coefficient (the nonzero coefficient of the monomial $\alpha^i$ with the maximum degree $i$) is integral, by Lemma \ref{Cauchy-bound}, all its roots have absolute values less than  $1+k^k\beta_{\max}^k$ for any ${\cal S}$. The required result then follows immediately.
\end{proof}

\begin{definition}[Linear degeneracy problem, \cite{Khachiyan_1995}]\label{linear-degeneracy} Given an $n\times k$ ($k<n$) matrix $W$, the linear degeneracy problem is to determine whether there exist $k$ rows of $W$ that are linearly dependent, i.e., whether a set ${\cal S}\subseteq [n]$ with $|{\cal S}|=k$ exists such that $\det W_{{\cal S}}=0$.		
\end{definition}

The linear degeneracy problem is NP-complete even when $W$ in Definition \ref{linear-degeneracy} are restricted to be integer matrices \cite{Khachiyan_1995}.
\begin{theorem}
	Both Problems 1 and 2 are NP-hard.
\end{theorem}
\begin{proof}
	As analyzed above, it suffices to prove the NP-hardness of Problem 2 with a scalar functional privacy. In this case, $F$ becomes a row vector $f$. We shall present a reduction from the linear degeneracy problem to Problem 2.
	
	Let $W={\bf col}\{w_i|_{i=1}^n\}$ be an $n\times k$ ($k<n$) integer matrix with full column rank, with $w_i\in {\mathbb R}^{1\times k}$. Denote the maximum absolute value of entries in $W$ by $\beta_{\max}$. Let $W^\bot$ be an $n\times (n-k)$ integer matrix with full column rank such that $(W^\bot)^\intercal W=0$. Such a $W^\bot$ can be determined in polynomial time as shown in \citep[Coros.3.2d]{schrijver1998theory}, and each entry of $W^\bot$ can have a size (the number of bits needed to express a number in binary format, also called encoding length) polynomially bounded by $k$ and $\beta_{\max}$. Let $\beta'_{\max}=\max\{|W^\bot_{ij}|\}$.
	
	Define an $n\times n$ matrix as
	$$H(\eta)=[W,W^\bot+\eta{\bf 1}_{n\times (n-k)}].$$
	Since $(W^\bot)^\intercal$ forms a basis for the left null space of $W$, it follows that $H(0)$ is non-singular. Note the coefficient matrix of $\eta$ in $H(\eta)$ is of rank one. As a result, the polynomial $\det H(\eta)$ in $\eta$  has a degree of at most one. As $\det H(\eta)$ is not identically zero ($\det H(0)\ne 0$), $\det H(\eta)$ has at most one zero for $\eta$. Observing this, let $\eta^*\in \{\beta_{\max}'+1,\beta_{\max}'+2\}$ be such that makes $\det H(\eta^*)\ne 0$. To determine $\eta^*$, one just needs to let $\eta=\beta_{\max}'+1$ and $\eta =\beta_{\max}'+2$ and then check whether $\det H(\eta)\ne 0$. Since each entry of $H(\eta)$ for $\eta=\beta_{\max}'+1$ or $\beta_{\max}'+2$ has a size polynomially bounded by $k$ and $\beta_{\max}$, so does the corresponding $\det H(\eta)$ \cite[Coro 3.3a]{schrijver1998theory}. In addition, $\det H(\eta)$ can be computed in polynomial time. Therefore, $\eta^*$ can be determined in polynomial time.
	
	Let $P=H(\eta^*)$, $\Gamma={\bf diag}\{I_k,2,3,...,n-k+1\}$, and $\alpha=1+k^k\beta^k_{\max}$. We construct the system parameters $(A,f)$ as
	$$A=P\Gamma P^{-1}, f=[\alpha,\alpha^2,\cdots,\alpha^n].$$ The maximum entry of $f$ is of size $\log_2\alpha^n=n\log_2(1+k^k\beta_{\max}^k)=n(k\log_2k+k\log_2\beta_{\max}+o(1))$, which is polynomially bounded by $k$ and $n$.\footnote{The little-o notation $f(n) = o(g(n))$ means $f(n)$ becomes insignificant relative to $g(n)$ as $n$ approaches infinity.} Moreover, $P^{-1}$ can be determined in polynomial time and is of the size polynomially bounded by $k$ and $n$ \cite[Coro 3.3a]{schrijver1998theory}. Therefore, the parameters $A,f$ can be constructed in polynomial time and have polynomially bounded sizes.

	We shall prove that the optimal value of Problem 2 associated with $(A,f)$ is no more than $n-k$, if and only if there exists a $k\times k$ singular submatrix in $W$. Let us keep in mind that from the construction, $W$ consists of $k$ linearly independent eigenvectors of $A$	associated with the eigenvalue $1$. The $(k+i)th$ column of $P$, given by $P_{k+i}$, is the eigenvector of $A$ associated with the eigenvalue $i+1$, $i=1,...,n-k$.
	
	For the one direction, suppose there is a $k\times k$ singular submatrix in $W$, whose rows are indexed by ${\cal S}_1\subseteq [n]$. Since $W$ has full column rank, by the property of matrix rank, there must be another set ${\cal S}_2\subseteq [n]$ with $|{\cal S}_2|=k$ such that ${\rm rank}W_{{\cal S}_2}=k-1$.\footnote{\label{footnote-rank} More precisely, this is due to the property that, for two sets ${\cal S}_3,{\cal S}_4\subseteq [n]$ such that ${\rm rank}W_{{\cal S}_3}=|{\cal S}_3|$, ${\rm rank}W_{{\cal S}_4}=|{\cal S}_4|$, and $|{\cal S}_4|>|{\cal S}_3|$, there must be some $i\in {\cal S}_4\backslash {\cal S}_3$ satisfying ${\rm rank}W_{\{i\}\cup{\cal S}_3}=|{\cal S}_3|+1$; see \citep[Page 73]{Murota_Book}.  Let ${\cal S}_0\subseteq [n]$ be such that ${\rm rank}M_{{\cal S}_0}=|{\cal S}_0|=k$, and suppose ${\rm rank}M_{{\cal S}_1}=k-k_0$, where $k_0>1$. Using this property, we know there exists $i\in {\cal S}_0\backslash {\cal S}_1$ such that ${\rm rank} M_{{\cal S}_1\cup \{i\}}=k-k_0+1$. Repeat this process $k_0-1$ times, and we get that there exists ${\cal S}'_0\subseteq {\cal S}_0\backslash {\cal S}_1$ making ${\rm rank} M_{{\cal S}_1\cup {\cal S}'_0}=k-1$. Hence, a set ${\cal S}_2\subseteq {\cal S}_1\cup {\cal S}'_0$ exists satisfying the proposed requirement.} Consider the eigenvalue $\lambda=1$. Since $W$ is a basis matrix of the null space of $A-I_n$ and ${\rm rank}W_{{\cal S}_2}<k$, it follows from Lemma \ref{composity} that ${\bf col}\{A-I_n,I_{{\cal S}_2}\}$ is of column rank deficient. On the other hand,  Lemma \ref{f-construct} yields
	$${\rm rank}\ {\bf col}\{fW,W_{{\cal S}_2}\}=k,$$noting that a set ${\cal S}\subseteq {\cal S}_2$ must exist satisfying ${\rm rank}W_{\cal S}=k-1=|{\cal S}|$. Again from Lemma \ref{composity}, ${\bf col}\{A-I,I_{{\cal S}_2},f\}$ has full column rank. As a result,
	$${\rm rank}\ {\bf col}\{A-I,I_{{\cal S}_2},f\}=n> {\rm rank}\ {\bf col}\{A-I,I_{{\cal S}_2}\},$$
	which means the system $(A,I_{{\cal S}_2},f)$ is not functionally observable from Lemma \ref{functional-theorem}. Consequently, upon letting ${\cal S}^*=[n]\backslash {\cal S}_2$, ${\cal S}^*$ is a feasible solution to Problem 2 associated with $(A,f)$. Since $|{\cal S}^*|=n-k$, the optimal value of Problem 2 is no more than $n-k$.
	
	For the other direction, suppose ${\cal S}^*\subseteq [n]$ is a feasible solution to Problem 1 associated with $(A,f)$ such that $|{\cal S}^*|\le n-k$. Let $\bar {\cal S}^*=[n]\backslash {\cal S}^*$. Notice that for each eigenvalue $\lambda_i=i+1$ of $A$, $i=1,...,n-k$, all entries in its eigenvector $P_{k+i}$ are positive since $\eta^*>\beta'_{\max}$. Therefore, to make ${\bf col}\{A-\lambda_iI,I_{\bar {\cal S}^*}\}$ column rank deficient, which is equivalent to making $[P_{k+i}]_{\bar {\cal S}^*}=0$ by Lemma \ref{composity}, it must hold ${\cal S}^*=[n]$. Hence, the only case to make $(A,I_{\bar {\cal S}^*},f)$ fail to be functionally observable while $|{\cal S}^*|\le n-k$ is that the following relation should hold
	$$n\ge {\rm rank}\ {\bf col}\{A-I,I_{\bar {\cal S}^*},f\}> {\rm rank}\ {\bf col}\{A-I,I_{\bar {\cal S}^*}\}.$$
	This requires ${\rm rank}\ {\bf col}\{A-I,I_{\bar {\cal S}^*}\}<n$, which is equivalent to ${\rm rank} W_{\bar {\cal S}^*}<k$ from Lemma \ref{composity}. Since $|\bar {\cal S}^*|=n-|{\cal S}^*|\ge k$, it follows that every $k\times k$ submatrix of $W_{\bar {\cal S}^*}$ must be singular, indicating  the linear degeneracy problem on matrix $W$ is yes.
	
	The above analysis indicates that verifying whether the optimal solution to Problem 2 on $(A,f)$ is no more than $n-k$ is equivalent to the linear degeneracy problem on $W$. Since the latter problem is NP-complete and all the reductions are in polynomial time and the involved parameters have polynomially bounded sizes, we conclude that Problem 2 is NP-hard.
\end{proof}

\begin{remark}
	We remark that the NP-hardness of Problem 1 can also be obtained from \citep[Theo. 1]{zhang2018observability}, which established the NP-hardness of determining the minimum number of sensors whose removal can destroy system observability with dedicated sensors (i.e., $C=I_n$). This means Problem 1 is NP-hard with $F=I_n$, under which circumstance the functional observability collapses to the conventional observability.
\end{remark}

\section{Algorithms} \label{sec-V}
In this section, we give exact algorithms for Problem 1 under a reasonable assumption that the eigenvalue geometric multiplicities of $A$ are bounded by a constant. We also present a greedy algorithm for Problem 2.

The following assumption on the computational availability of eigenvalues and eigenvectors of $A$ is adopted.

\begin{assumption}\label{assumption-eigen}
	Suppose that the eigenvalues and eigenvectors of $A$ are computationally available. Moreover, suppose there are $q$ distinct eigenvalues in ${\rm eig}(A)$, the $i$th one denoted by $\lambda_i$, and $X_i\in {\mathbb C}^{n\times k_i}$ consists of the maximum number of linearly independent eigenvectors (i.e., $X_i$ is the eigenbasis) associated with $\lambda_i$.
\end{assumption}

\begin{remark}
	It is worth mentioning that, while eigenvalues and eigenvectors of a matrix can be computed to any prescribed precision in theory, the practical limitations of numerical methods and the conditioning of the matrix may sometimes make it difficult to achieve high precision in practice \cite{golub2013matrix}. On the other hand, the eigenvalues and eigenvectors of adjacency or Laplacian matrices of large-scale sparse graphs have been extensively studied \cite{mesbahi2010graph}, and there are specialized algorithms for computing eigenvalues of sparse matrices more efficiently and accurately than general-purpose algorithms \cite{saad2011numerical}. This makes Assumption \ref{assumption-eigen} reasonable for studying sparsely-connected networks.
\end{remark} 

\begin{assumption}\label{diagonal-assumption}
	The state matrix $A$ is diagonalizable.
\end{assumption}

\begin{remark} Equivalently, this assumption requires that $\sum_{i=1}^q k_i=n$.
	Diagonalizable matrices are quite common in system modeling and control. For example, all symmetric matrices, naturally arising in Laplacian matrices and adjacency matrices of undirected graphs, are diagonalizable. Moreover, the weighted Laplacian matrices of strongly connected directed graphs and adjacency matrices of random networks are mostly diagonalizable \cite{mesbahi2010graph,tao2017random}.
\end{remark}
\subsection{Algorithms for Problem 1}
According to Lemma \ref{functional-theorem}, under Assumption \ref{diagonal-assumption}, a natural idea to find a minimum set (i.e., a set with the minimum cardinality) ${\cal S}\subseteq [n]$ such that $(A,I^{[n]\backslash {\cal S}},F)$ fails to be functionally observable is to determine
the minimum set ${\cal S}_i$ for each eigenvalue $\lambda_i\in {\rm eig}(A)$ such that
\begin{equation} \label{rank-cond}
	{\rm rank}  \left[\begin{array}{c}
		A-\lambda_i I_n \\
		I^{[n]\backslash {\cal S}_i} \\
		F
	\end{array}\right]>{\rm rank}  \left[\begin{array}{c}
		A-\lambda_i I_n \\
		I^{[n]\backslash {\cal S}_i} \\
	\end{array}\right],
\end{equation}
and then find the minimum $|{\cal S}_i|$ over $i\in [q]$. In the following, we characterize the minimum set ${\cal S}_i$ that satisfies (\ref{rank-cond}).

\begin{proposition}\label{minimum-cond} Let ${\cal S}_i^*$ be a set with the minimum cardinality that satisfies (\ref{rank-cond}). Then, it holds that
	\begin{equation} \label{rank-cond-2a}
		{\rm rank}  \left[\begin{array}{c}
			A-\lambda_i I_n \\
			I^{[n]\backslash {\cal S}^*_i} \\
			F
		\end{array}\right]=n,\end{equation}		\begin{equation} \label{rank-cond-2b}  {\rm rank}  \left[\begin{array}{c}
			A-\lambda_i I_n \\
			I^{[n]\backslash {\cal S}^*_i} \\
		\end{array}\right]=n-1.
	\end{equation}Moreover, ${\cal S}_i^*$ (if exists) must be a {\emph{minimal set}} that satisfies (\ref{rank-cond-2b}) (by `minimal set' we mean ${\cal S}_i^*$ satisfies (\ref{rank-cond-2b}), but any proper subset $\bar {\cal S}_i^*\subseteq {\cal S}_i^*$ cannot satisfy (\ref{rank-cond-2b})). 	
\end{proposition}
\begin{proof}
	First, suppose that ${\cal S}_i^*$ makes
	\[
	{\rm rank}  \left[\begin{array}{c}
		A-\lambda_i I_n \\
		I^{[n]\backslash {\cal S}^*_i} \\
		F
	\end{array}\right]=n-k_0>{\rm rank}  \left[\begin{array}{c}
		A-\lambda_i I_n \\
		I^{[n]\backslash {\cal S}^*_i} \\
	\end{array}\right]=n-k_0',\]where $k_0'>k_0>0$. Since ${\bf col}\{A-\lambda_iI_n,I,F\}$ has full column rank, there exists ${\cal S}_a\subseteq {\cal S}_i^*$ with $|{\cal S}_a|=k_0$, such that ${\rm rank}\,{\bf col}\{A-\lambda_iI,I^{([n]\backslash {\cal S}_i^*)\cup {\cal S}_a},F\}=n$ (see footnote \ref{footnote-rank}). Notice that ${\rm rank}\,{\bf col}\{A-\lambda_iI,I^{([n]\backslash {\cal S}_i^*)\cup {\cal S}_a}\}\le n-k_0'+k_0<n$. Therefore, ${\cal S}_i^*\backslash {\cal S}_a$ also satisfies (\ref{rank-cond}). However, $|{\cal S}_i^*\backslash {\cal S}_a|<|{\cal S}_i^*|$, contradicting the optimality of ${\cal S}_i^*$. Hence, (\ref{rank-cond-2a}) must be satisfied.
	
	Now suppose ${\cal S}_i^*$ satisfies
	\[
	{\rm rank}  \left[\begin{array}{c}
		A-\lambda_i I_n \\
		I^{[n]\backslash {\cal S}^*_i} \\
		F
	\end{array}\right]=n>{\rm rank}  \left[\begin{array}{c}
		A-\lambda_i I_n \\
		I^{[n]\backslash {\cal S}^*_i} \\
	\end{array}\right]=n-k_0',\]where $k_0'\ge 2$. Since ${\rm rank}\,{\bf col}\{A-\lambda_iI,I\}=n$, again from the same reasoning in footnote \ref{footnote-rank}, there exists ${\cal S}_a\subseteq {\cal S}_i^*$ such that ${\rm rank}\,{\bf col}\{A-\lambda_iI,I^{([n]\backslash {\cal S}_i^*)\cup {\cal S}_a}\}=n-1$. Notice that ${\rm rank}\,{\bf col}\{A-\lambda_iI,I^{([n]\backslash {\cal S}_i^*)\cup {\cal S}_a},F\}\ge {\rm rank}\,{\bf col}\{A-\lambda_iI,I^{[n]\backslash {\cal S}_i^*},F\}=n$. This indicates ${\cal S}_i^*\backslash {\cal S}_a$ also satisfies (\ref{rank-cond}), still contradicting the optimality of ${\cal S}_i^*$. Hence, (\ref{rank-cond-2b}) must be satisfied.
	
	If ${\cal S}_i^*$ exists, it must satisfy (\ref{rank-cond-2a}) and (\ref{rank-cond-2b}) simultaneously. If ${\cal S}_i^*$ is not a minimal set satisfying (\ref{rank-cond-2b}), by definition, one can remove some elements $\Delta$ from ${\cal S}_i^*$ so that (\ref{rank-cond-2b}) is maintained. Meanwhile, (\ref{rank-cond-2a}) still holds since adding more rows cannot reduce the rank. A new set ${\cal S}_i^*\backslash \Delta$ is obtained that satisfies (\ref{rank-cond}), contradicting the optimality of ${\cal S}_i^*$.
\end{proof}

By virtue of Proposition \ref{minimum-cond}, to determine ${\cal S}_i^*$ for each $\lambda_i\in {\rm eig}(A)$, one can determine all minimal sets that satisfy (\ref{rank-cond-2b}), and then find out those that satisfy (\ref{rank-cond-2a}). Then, ${\cal S}_i^*$ is such a set with the minimum cardinality. Note that ${\cal S}_i^*$ may be empty for some $\lambda_i\in {\rm eig}(A)$. But with Assumption \ref{diagonal-assumption}, there is at least one $\lambda_i\in {\rm eig}(A)$ such that ${\cal S}_i^*$ is not empty. Indeed, the worst-case solution to Problem 1 is the full state set $[n]$.
After determining ${\cal S}_i^*$ for each $\lambda_i\in {\rm eig}(A)$, the optimal solution is the set ${\cal S}_i^*$ with the minimum cardinality over $i\in [q]$. %

\subsubsection{Simple dynamic case}
We first consider the case where $A$ has no repeated eigenvalues, i.e., the simple dynamic case. In this situation, $k_i=1\ \forall i$ and $q=n$. If a set ${\cal S}_i$ satisfies (\ref{rank-cond-2b}), by Lemma \ref{composity}, it holds that
$$[X_i]_{[n]\backslash {\cal{S}}_i}=0.$$
Hence, the minimal set satisfying (\ref{rank-cond-2b}) is unique, which is the support of $X_i$, given by
$$\bar {\cal S}_i^*={\rm supp} X_i\doteq \{j\in [n]: [X_i]_j\ne 0\}. $$
Let ${\cal S}_i^*$ be such $\bar {\cal S}_i^*$ that satisfies (\ref{rank-cond-2a}). For ease of description, if $\bar {\cal S}_i^*$ does not satisfy (\ref{rank-cond-2a}), we assign ${\cal S}_i^*=[n]$. As a result, the optimal solution to Problem 1 (denoted by ${\cal S}^*_{P_1}$) is
$${\cal S}^*_{P_1}=\arg\nolimits_{i\in [n]} \min |{\cal S}_i^*|. $$
\subsubsection{Bounded eigenvalue geometric multiplicity case}We now generalize the simple dynamic case to systems with bounded eigenvalue geometric multiplicities. More precisely, we consider systems satisfying the following assumption:
\begin{assumption} \label{bound-assumption}
	The geometric multiplicities of eigenvalues of $A$ are bounded by some fixed constant $k_c$, i.e., $k_i\le k_c$  $\forall i\in [q]$ as $n$ increases.
\end{assumption}

The above assumption can be satisfied by most practical systems. The simple dynamic case is one such with $k_c=1$. Besides, when modeling networks of coupled oscillators, power grids, diffusively couple networks, epidemiological networks using graphs, a common setting is that each node has a self-loop \cite{mesbahi2010graph,zhang2021structural,LucaSynchronization,ye2018evolution}, under which these networks can be controllable using some constant number of inputs regardless of the network size \cite{plos_one,A.Ol2014Minimal}. This indicates the above assumption is satisfied for these networks, since the minimum number of inputs for achieving controllability equals the maximum eigenvalue geometric multiplicities of system state matrices \cite{Y_Zhang_2018}.

From the previous analysis, the key step for solving Problem 1 is to determine the collection ${\Omega_i}$ of all minimal sets that satisfy (\ref{rank-cond-2b}) for a given $\lambda_i\in {\rm eig}(A)$. We provide Algorithm \ref{alg1} for this purpose. In this algorithm, ${\cal W}_j'$ is a {\emph{maximal} set satisfying ${\rm rank}[X_i]_{{\cal W}_j'}=k_i-1$, meaning that adding additional rows to $[X_i]_{{\cal W}_j'}$ will increase its rank. Therefore, by Lemma \ref{composity}, $[n]\backslash {\cal W}_j'$ is a minimal set satisfying (\ref{rank-cond-2b}). With Assumption \ref{bound-assumption}, $|\Omega_i|\le {\tiny{\left(\begin{array}{l} n\\k_i \end{array}\right)}} \le n^{k_c}$. After the determination of $\Omega_i$ for $i\in [q]$, the rest is similar to the simple dynamic case. We collect the whole procedure in Algorithm \ref{alg2}, and state the following result.
	
	\begin{theorem} \label{main-theo}
		Under Assumptions 1-3, Algorithm \ref{alg2} is able to find an optimal solution to Problem 1 in time $O(n^{k_c+2}k_c^3)$.
	\end{theorem}
	\begin{proof}
		The reason why Algorithm \ref{alg2} returns an optimal solution has been explained in Proposition \ref{minimum-cond} and the main contexts. Here we just need to justify the computational complexity. Step 1 incurs $O(n^3)$ time \cite{golub2013matrix}. For each eigenbasis $X_i$, checking whether its $k_i-1$ rows are linearly independent can use the singular value decomposition (SVD), which takes $O(k_i^3)$ time \cite{golub2013matrix}. For each ${\cal W}_j$ in Algorithm \ref{alg1}, it takes $O(nk_i^3)$ time to get ${\cal W}_j'$ via SVD. For each member ${\cal S}_{ij}\in {\Omega_i}$, it takes $O(k_i^2(n+r))$ time to get ${\bar {\cal S}}_{ij}$. Since $|\Omega_i|\le n^{k_i}$, determining ${\cal S}_i^*$ incurs time $O(n^{{k_i}+1}k_i^3)$. As $q\le n$ and $k_i\le k_c$, the total time complexity is at most $O(n^{k_c+2}k_c^3)$.
	\end{proof}
	
	\begin{remark}
		Theorem \ref{main-theo} makes it clear that the computation cost of Algorithm \ref{alg2} scales exponentially with $k_c$. This indicates it is the eigenvalue geometric multiplicities of $A$ that cause the computational intractability of Problem 1.
	\end{remark}
	
	\begin{algorithm} 
		{\small {{{{
							\caption{: Enumerating all minimal sets satisfying (\ref{rank-cond-2b})} 
							\label{alg1} 
							\begin{algorithmic}[1] 
								\REQUIRE The eigenbasis $X_i$ of $\lambda_i\in {\rm eig}(A)$  
								\ENSURE  The collection ${\Omega_i}$ of the minimal sets satisfying (\ref{rank-cond-2b})  
								\STATE  Determine all sets $\{{\cal W}_j\}_{j=1}^{N_1}$ that contain $k_i-1$ linearly independent rows of $X_i$ ($N_1$ is the number of such sets).
								\FOR{$j =1$ to $N_1$}
								\STATE ${\cal W}_j' =   {\cal W}_j \cup \left\{k:{\rm rank}[X_i]_{{\cal W}_j\cup \{k\}}=k_i-1 \right\}$	
								\STATE ${\cal S}_{ij}=[n]\backslash {\cal W}_j'$
								\ENDFOR
								\STATE Return ${\Omega_i}=\{{\cal S}_{ij}\}_{j=1}^{N_1}$.
					\end{algorithmic}}}
		}}}
	\end{algorithm}
	
	\begin{algorithm} 
		{\small {{{{
							\caption{: Algorithm for Problem 1} 
							\label{alg2} 
							\begin{algorithmic}[1] 
								\REQUIRE Parameters $(A,F)$ satisfying Assumption \ref{assumption-eigen}-\ref{bound-assumption}  
								\ENSURE  The optimal solution ${\cal S}_{P_1}^*$ to Problem 1  
								\STATE Calculate the eigenbases $\{X_i|_{i=1}^q\}$ of $A$.
								\FOR{$i =1$ to $q$}
								\STATE  Determine the collection ${\Omega_i}$ of the minimal sets satisfying (\ref{rank-cond-2b}) using Algorithm \ref{alg1}.
								\STATE  For each member ${\cal S}_{ij}\in \Omega_i$, check whether it satisfies (\ref{rank-cond-2a}), if yes, let $\bar {\cal S}_{ij}={\cal S}_{ij}$; otherwise, let $\bar {\cal S}_{ij}=[n]$. Find ${\cal S}_i^*=\arg \min |\bar {\cal S}_{ij}|$.
								\ENDFOR
								\STATE Return ${\cal S}_{P_1}^*=\arg\nolimits_{i\in [q]} \min |{\cal S}_i^*|$.
					\end{algorithmic}}}
		}}}
	\end{algorithm}
	
	\subsection{Greedy algorithm for Problem 2}
	By the definition of functional observability, if we obtain the optimal solution ${\cal S}^*_{P_1i}$ to Problem 1 associated with $(A,f_i)$ for each row $f_i$ of $F$, then $\bigcup_{i=1}^r {\cal S}^*_{P_1i}$ is a feasible solution to Problem 2. However, such a solution ignores the possible overlaps among different $f_i's$, which may be far from the optimal one. In what follows, we provide a greedy algorithm for Problem 2. This algorithm is based on the following result, which generalizes Proposition~\ref{minimum-cond}.
	
	\begin{proposition} \label{alg-base}
		With Assumption \ref{diagonal-assumption}, suppose for one $\lambda_i\in {\rm eig}(A)$, some ${\cal T}_k\subseteq [n]$, and the $j$th row $f_j$ of $F$, it holds
		\begin{equation}\label{equal-condi}{\rm rank}  \left[\begin{array}{c}
				A-\lambda_i I_n \\
				I^{{\cal T}_k} \\
				f_{j}
			\end{array}\right]={\rm rank}  \left[\begin{array}{c}
				A-\lambda_i I_n \\
				I^{{\cal T}_k}
			\end{array}\right]=n_{ij}\le n.\end{equation}
		If $\Delta^*_{kj}\subseteq {\cal T}_k$ is a set with the minimum cardinality satisfying
		\begin{equation} \label{functional-cond-2} {\rm rank}  \left[\begin{array}{c}
				A-\lambda_i I_n \\
				I^{{\cal T}_k\backslash \Delta^*_{kj}} \\
				f_{j}
			\end{array}\right]>{\rm rank}  \left[\begin{array}{c}
				A-\lambda_i I_n \\
				I^{{\cal T}_k\backslash \Delta^*_{kj}}
			\end{array}\right],\end{equation}it must hold that
		\begin{equation}\label{greedy-relation}
			{\rm rank}  \left[\begin{array}{c}
				A-\lambda_i I_n \\
				I^{{\cal T}_k\backslash \Delta^*_{kj}} \\
				f_{j}
			\end{array}\right]=n_{ij}, {\rm rank}  \left[\begin{array}{c}
				A-\lambda_i I_n \\
				I^{{\cal T}_k\backslash \Delta^*_{kj}}
			\end{array}\right]=n_{ij}-1,
		\end{equation}
		and $\Delta^*_{kj}$ is a minimal set satisfying the second equality of (\ref{greedy-relation}).
	\end{proposition}
	\begin{proof}
		The proof is similar to that of Proposition \ref{minimum-cond}, thus omitted.
	\end{proof}
	
	It is clear that if ${\cal T}_k=[n]$, $n_{ij}=n$, then Proposition \ref{alg-base} collapses to Proposition \ref{minimum-cond}. As a result, when (\ref{equal-condi}) holds for every $\lambda_i\in {\rm eig}(A)$, we can use a similar manner to Algorithm \ref{alg2} to find
	a set $\Delta^*_{kj}$ with the minimum cardinality such that (\ref{functional-cond-2}) is true for some $\lambda_i\in {\rm eig}(A)$. Denote such a process by $\Delta^*_{kj}\leftarrow {\rm Alg2}[A,f_j,{\cal T}_k]$. That is, ${\rm Alg2}[A,f_j,{\cal T}_k]$ finds the minimum set $\Delta^*_{kj}$ from ${\cal T}_k$ such that $(A,I^{{\cal T}_k\backslash \Delta^*_{kj}},f_j)$ becomes functionally unobservable ($\Delta_{kj}^*$ will be empty if $(A,I^{{\cal T}_k},f_j)$ is already functionally unobservable). We formulate the greedy algorithm for Problem 2 as Algorithm \ref{alg3}.
	
	\begin{algorithm} 
		{\small {{{{
							\caption{: Greedy Algorithm for Problem 2} 
							\label{alg3} 
							\begin{algorithmic}[1] 
								\REQUIRE Parameters $(A,F)$ satisfying Assumption \ref{assumption-eigen}-\ref{bound-assumption}  
								\STATE Calculate the eigenbases $\{X_i|_{i=1}^q\}$ of $A$.
								\STATE Initialize ${\cal T}_0=[n],{\cal F}=[r],k=0$
								\WHILE{$|{\cal T}_k|>0$ and $|{\cal F}|>0$}
								\FOR{$j\in {\cal F}$}
								\STATE $\Delta^*_{kj}\leftarrow {\rm Alg2}[A,f_j,{\cal T}_k]$.
								\ENDFOR
								\STATE $\Delta^*_{kj^*}\leftarrow \arg \min_{j\in {\cal F}} |\Delta^*_{kj}|$
								\STATE Update ${\cal T}_{k+1}\leftarrow {\cal T}_k\backslash \Delta^*_{kj^*}$,
								${\cal F}\leftarrow {\cal F}\backslash \{j^*\}$, and $k\leftarrow k+1$.
								\ENDWHILE
								\STATE Return a solution $[n]\backslash {\cal T}_k$

					\end{algorithmic}}}
		}}}
	\end{algorithm}
	
	Compared to the naive method mentioned at the beginning of this subsection, the advantage of Algorithm \ref{alg3} lies in that, it
	not only guarantees to protect one scalar functional privacy $f_ix(t)$ per step but also accounts for the relations between different $f_i$'s. 
	The computation time of Algorithm \ref{alg3} is dominated by the subroutine ${\rm Alg2}[A,f_j,{\cal T}_k]$, which runs at most $r^2$ times. Therefore, the time complexity of Algorithm \ref{alg3} is
	$O(r^2n^{k_c+2}k_c^3)$. When $k_c$ is large, this is a huge computational burden.

	\begin{remark}
		Based on Proposition \ref{alg-base}, both Algorithms \ref{alg2} and \ref{alg3} can be trivially extended to the case where the adversarial observers or eavesdroppers have access only to a restricted set of full states. In this case, we just need to change the full state set $[n]$ to the aforementioned restricted set (${\cal T}_k$ alike).
	\end{remark}
	
	\section{Illustrative Example} \label{sec-VI}
	Consider a network system with {\small
		$$A= \left[\begin{array}{cccccc}
			1&0&0&0&0&0\\
			3&5&2&0&0&0\\
			4&0&4&0&0&0\\
			2&0&0&2&0&0\\
			0&2&1&3&6&0\\
			0&0&0&5&4&9
		\end{array}\right],$$}whose topology is given in Fig. \ref{examp-graph}. It turns out that ${\rm eig}(A)=\{1,5,4,2,6,9\}$, and the corresponding eigenvectors have supports respectively as $\bar {\cal S}_1^*=\{1,\cdots,6\}$, $\bar {\cal S}_2^*=\{2,5,6\}$, $\bar {\cal S}_3^*=\{2,3,5,6\}$, $\bar {\cal S}_4^*=\{4,5,6\}$, $\bar {\cal S}_5^*=\{5,6\}$, and $\bar {\cal S}_6^*=\{6\}$. 
	
	First, let us consider Problem 1 with $F_1=I_6$. Algorithm \ref{alg2} returns a solution $\{6\}$. This means blocking state $x_6$ only is enough to protect the full state vector-wisely. Now, let $F_2=[0,1,1,1,0,0]/3$, meaning that the functional privacy is the average state of cluster $\{x_2,x_3,x_4\}$. Two solutions, $\{2,5,6\}$ or $\{4,5,6\}$, are found via Algorithm \ref{alg2}. It can be validated that both are optimal.

	Next, consider $F_3={\bf col}\{e^\intercal_3,e^\intercal_4,e^\intercal_5\}$ ($n=6$), i.e., protecting states $\{x_3,x_4,x_5\}$. Using Algorithm \ref{alg2}, we get an optimal solution with $2$ states $\{5,6\}$, implying that at least two states need to be blocked. Comparing these solutions, it turns out that although blocking state $x_6$ is enough to protect the full state vector-wisely, it cannot protect states $\{x_3,x_4,x_5\}$. Finally, suppose we are to protect state variables $\{x_3,x_4,x_5\}$ entry-wisely. Implementing Algorithm \ref{alg3} on $(A,F_3)$, we get ${\cal T}_1=\{1,2,3,4\}$ and ${\cal F}_1=\{3,4\}$, ${\cal T}_2=\{1,2,3\}$ and ${\cal F}_2=\{3\}$, and ${\cal T}_3=\{1\}$ and ${\cal F}_3=\emptyset$, which means the solution is $\{2,3,4,5,6\}$. By exhaustive search, it can be validated that this solution is optimal. 
	
	This above example shows that the minimum set of states needed to be blocked varies drastically with the functional privacy to be protected; and even for the same functional privacy, the vector-wise protection and entry-wise protection can lead to
	drastically different solutions. 
	\begin{figure}
		\centering
		\includegraphics[width=1.5in]{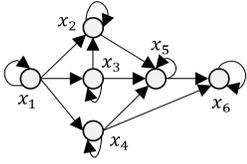}\\
		\caption{Network topology of the system in Section \ref{sec-VI}.} \label{examp-graph}
	\end{figure}
	
	\section{Conclusion} \label{sec-VII}
	This paper addressed and investigated the problem of protecting functional privacy in a network by blocking the minimum set of state variables from direct measurements. It has been proven that both the vector-wise and entry-wise functional privacy protection problems are NP-hard, but an exact algorithm with polynomial time complexity is presented for the vector-wise problem by assuming a reasonable constant bound on the system eigenvalue geometric multiplicities. A greedy algorithm for the entry-wise problem is further provided. Our approach is based on functional observability and a PBH-like criterion, and the effectiveness of the proposed approach is demonstrated through an example. In the future, we plan to extend our study to a structured system model.

	\bibliographystyle{elsarticle-num}
	{\footnotesize
		\bibliography{yuanz3}
	}

\end{document}